\documentstyle[preprint,prd,aps]{revtex}

\begin{document}


\draft


\title{ 
 On the Stability of the\\
GPS Magnetic Monopole Solution}

\author{Bj\o rn Jensen
\footnote{e-mail address: bjensen@boson.uio.no}}

\address{Institute of Physics,
University of Oslo,\\
Box 275, N-0314 Blindern, Oslo 3,\\
Norway}

\author{Parviz H. Mani
\footnote{e-mail address: parviz@vanosf.physto.se}}

\address{Institute of Theoretical Physics,
University of Stockholm,\\
Box 6730, Vanadisv\"agen 9, S-113 85 Stockholm,\\
Sweden}


\maketitle

\begin{abstract}
The question of
the stability of the four dimensional 
Gross-Perry-Sorkin Kaluza-Klein magnetic
monopole solution is investigated within the
framework of a N=2, D=5 supergravity theory.
We show that this solution does not support 
a spin structure of the Killing type and is 
therefore, contrary to previous expectations, not 
necessarily stable.
\end{abstract}



 \begin{center}{\it OSLO-TP 3-97,
 USITP 97-01}\end{center}

\newpage




The recent huge interest in solitonic solutions in various
supergravity (SUGR) theories is partially due to the
fundamental role they play in string and 
M-theory. There fundamental strings/branes 
are exchanged with non-singular solitonic strings/branes
under S-duality.
In this note we will focus on 
the fundamental question of  
whether
 the solitonic
Gross-Perry-Sorkin 
($\equiv$GPS) Kaluza-Klein (KK) magnetic monopole
solution \cite{Gross,Sorkin1} in four dimensions
 is a stable one, or not. This solution is
of intrinsic interest to the question of S-duality in  
four dimensions \cite{Gibbons}. 

The stability of a gravitating configuration is in general a difficult
task to prove. However, in theories which allow for the presence of 
extended local supersymmetry (SUSY) this task is almost trivialized 
in certain cases. This is so since these theories in general carry charges
of both the electric ($Q_E$), as well as of the magnetic $(Q_M$) type.
These charges put a lower bound (of the Bogomol'nyi type) on the value of the mass $M$ of the
configuration, which in certain appropriate units always
can be written in the form
\begin{equation}
M\geq\sqrt{\sum_{i}Q_{iE}^2+\sum_{i}Q_{iM}^2}
\end{equation}
The sums run over all the available charge degrees of freedom of the system.
When this relation is saturated (with some $Q_j\neq 0$), it follows that the configuration
is part of a short super-multiplet, with half of the available supersymmetries
being broken. 
Solitonic states of this kind correspond to the BPS states.
Hence, the extremal value of the
mass, which is attained at saturation, is not only
protected from below, i.e. the configuration can not radiate energy
to infinity, but it is also protected from acquiring more mass,
since if it did it would also leave the short representation.
Hence, these states are stable, also quantum mechanically.
The question of the stability of the GPS monopole has also been investigated
within a classical KK framework
in \cite{Sorkin2} where techniques which are imported from SUGR were employed. We will
investigate the same problem within the {\em consistent} framework provided by
N=2, D=5 SUGR theory.

This work is organized as follows. 
We will first very briefly consider the minimal N=2, D=5 SUGR
theory. This theory can be
obtained via dimensional reduction from simple eleven dimensional SUGR,
which is believed to be related to the low energy limit of M-theory.
We then briefly remind ourselves of a derivation,
along the lines of Witten and Nester \cite{Witten,Nester}, 
of a Bogomol'nyi type bound for five dimensional classical KK-theory. 
This first part of this work is only meant to provide a
sufficient and, most importantly, a consistent background for our
reasoning. 
We next consider the Killing spinor equation, and we show that
the GPS monopole solution can not saturate 
the Bogomol'nyi bound, since it can not
support a spin structure of the Killing type.
Contrary to previous expectations \cite{Sorkin2},
this implies that the 
magnetic monopole solution is not
necessarily stable.  
   

We will consider the usual N=0, D=5 KK-theory
as being embedded into ``minimal'' N=2, D=5 SUGR
theory. This theory,
when no five dimensional matter multiplets are included, and
when we assume that the ``cylinder'' conditions hold,
will reduce to a N=2, D=4 Einstein-Maxwell-dilaton theory
upon KK dimensional reduction.
This theory will
reduce to a bosonic KK-theory in five
and four dimensions when the gravitini contribution is also 
consistently neglected.
However, it is clear that the solution space of this truncated 
theory is
only a very restricted
subset of the corresponding
solution space of standard five dimensional  
KK-theory. This is due to the
severe restrictions 
which are implied
on the allowed field configurations 
by the constraint which secures that the gravitini always vanish.
The full N=2, D=5 SUGR lagrangian
density can conveniently be written in the form 
(see e.g. E. Cremmer in\cite{Hawking})
\begin{eqnarray}
E^{-1}{\cal L}=&-&\frac{1}{4}R-\frac{i}{2}\overline{\Psi}^a_\mu
\gamma^{\mu\nu\rho}D_\nu (\frac{\omega +
\hat{\omega}}{2})\Psi_{\rho a}
-\frac{1}{4}\hat{F}_{\mu\nu}\hat{F}_{\rho\sigma}g^{\mu\rho}
g^{\nu\sigma}+\nonumber\\
&+&\frac{E^{-1}}{6\sqrt{3}}\epsilon^{\mu\nu\rho\sigma\lambda}\hat{F}
_{\mu\nu}\hat{F}_{\rho\sigma}\hat{A}_\lambda 
-\frac{i\sqrt{3}}{16}(\hat{F}_{\mu\nu}+\tilde{F}_{\mu\nu}
\overline{\Psi}^{\rho c}\gamma_{[\rho}
\gamma^{\mu\nu}\gamma_{\sigma ]}\Psi^\sigma\, _c)
\end{eqnarray}
where
$\tilde{F}_{\mu\nu}\equiv \hat{F}_{\mu\nu}+
\frac{\sqrt{3}}{4}\overline{\Psi}_\mu\, ^c\Psi_{\nu c}$.
$\hat{F}$ is the field strength of the five dimensional 
(abelian) ``photon'' $\hat{A}_\mu$ 
($\hat{F}=d\hat{A}$),
$\Psi_\mu\, ^a$ are the two gravitini (hence lower case Latin indices
at the beginning of the alphabet $\in\{ 1,2\}$),
$R$ is the D=5 Ricci curvature scalar, and $E$ is the
determinant of the f\"unf-bein field $E_\mu\, ^A$.
$\gamma^{\mu_1...\mu_n}$ denotes the totally skew product of n $\gamma$
matrices (to be defined below).
Capital Latin letters denote local f\"unf-bein indices, and lower case Latin
in the middle of the alphabet
denote global four indices, while
Greek ones denote global five indices. We will
use the ``mostly minus'' metric signature.
The f\"{u}nf-bein is given in terms of the vier-bein $\hat{E}^j$ by
\begin{equation}
E^j=e^{\frac{\phi}{2\sqrt{3}}}\hat{E}^j \,\, ,\,\, E^5=
E_\mu\, ^5dx^\mu=e^{-\frac{\phi}{\sqrt{3}}}(dx^5+A)
\end{equation}
$A=A_j(x^i)dx^j$ is the 
four-dimensional
U(1) gauge field (i.e. we will assume that 
the fifth dimension is  
topologically equivalent to $S^1$), with the corresponding field strength given
by $F=dA$, and $\phi$ is the four dimensional dilaton field. 
Variation of the action with
respect to the connection 
give us the expression for the (super) spin-connection  
\begin{eqnarray}
\hat{\omega}_{\mu RS}=\omega_{\mu RS}+\frac{i}{2}(\overline{\Psi}_\mu\, ^a\gamma_S\Psi_{Ra}
-\overline{\Psi}_R\, ^a\gamma_\mu\Psi_{Sa}+\overline{\Psi}_S\, ^a\gamma_R\Psi_{\mu a})
\end{eqnarray}
$\omega_{\mu RS}=\omega_{\mu RS}(E)$ 
is the purely gravitational spin-connection.
The action is invariant under the {\em two} 
local SUSY transformations given by
\begin{eqnarray}
&& \delta_\epsilon E_{\mu}\,^A=-i\overline{\epsilon}^a\gamma^A
\Psi_{\mu a}\nonumber \\
&& \delta_\epsilon A_\mu =-\frac{\sqrt{3}}{4}\overline{\epsilon}^a
\Psi_{\mu a}\nonumber\\
&& \delta_\epsilon\Psi_{\mu a}=[D_\mu ({\omega})+\frac{1}{4\sqrt{3}}
\tilde{F}_{\rho\sigma}
(\gamma^{\rho\sigma}\gamma_\mu +2\gamma^\rho\delta_\mu\, ^\sigma )]
\epsilon_a
\equiv \tilde{D}_\mu\epsilon_a
\end{eqnarray}
$D_\mu$ is the five dimensional
gravitational covariant derivative operator, and $\epsilon_a$ is taken
to be symplectic 
anti-commuting Majorana spinors. That is, $\epsilon_b=\Omega_b\, ^a\epsilon_a$
where $(\Omega_b\, ^a)$ is a symplectic structure, and each $\epsilon$ is a four spinor.
In five dimensions we can realize the Dirac algebra $\{\gamma^A,\gamma^B\} =2\eta^{AB}$
by the usual $4\times 4$ Dirac matrices in four dimensions supplemented by
the usual fifth matrix $\gamma^4\equiv i\gamma^0\gamma^1\gamma^2\gamma^3$.
The Dirac conjugate $\overline{\epsilon}$ of a spinor $\epsilon$ is as usual
given by $\overline{\epsilon}=\epsilon^\dagger\gamma^0$.

In order to obtain the five dimensional bosonic KK-theory we must
reduce the above theory. One way to
do this is to set the gravitini $\Psi_{a\mu}$,
as well as the five dimensional
vector potential $\hat{A}_{\mu}$ to zero. From
the expression for the action we see that we formally are left
with five dimensional vacuum Einstein theory. This truncated theory is
consistent at the tree level, since no extra constraints on the fields arise from
the equations of motion, and provided that no fermions are
induced when a local SUSY transformation is performed.
A constraint of the last type is
in general satisfied provided that there exists 
at least one non-trivial
spinor field $\epsilon$ with the property that
$\tilde{D}_\mu\epsilon = 0$.
Any solution of the resulting theory is thus also invariant
under at least one local SUSY transformation, and they
are in this sense supersymmetric bosonic configurations.  
 
The fact that BPS states actually saturate the 
lower bound on the energy of the system,
 is quite generally seen as follows
(we will assume that no event horizons are present, and that 
four dimensional space time
is topologically simply connected).
Introduce the Witten-Nester tensor $E^{\mu\nu}$ defined by \cite{Witten,Nester}
\begin{equation}
E^{\mu\nu}=\frac{1}{2}\overline{\epsilon}\gamma^{\mu\nu\rho}
\nabla_\rho\epsilon +c.c.
\end{equation}
$\epsilon$ is a spin structure, and
${\nabla}$ is some appropriately chosen
super-covariant derivative operator. In our case it is natural to identify $\nabla=
\tilde{D}$. This choice coincides with the one in \cite{Sorkin2}.
We will only consider pure Einstein theory in five dimensions, so that
the five dimensional ``photon'' must be set to vanish
 $\hat{F}_{\rho\sigma}=0$.
Hence,
$\tilde{D}_\rho$ simply reduces to the 
gravitational spin connection $D_\rho$ in five dimensions.
Consider now the integral of $D_\nu E^{\mu\nu}$ over some space like
three surface $\Sigma$ with an outer boundary $\partial\Sigma$ 
placed at ``spatial infinity''.
 With the help of Stoke's theorem this integral
can be written
\begin{eqnarray}
\int_{\partial\Sigma} dS_\mu D_\nu E^{\mu\nu}
&\equiv&\overline{\epsilon}_\infty
\gamma^\mu\epsilon_\infty  \, ^5P_\mu ^{ADM}\nonumber\\
&\equiv& {\cal M}
\end{eqnarray}
where we have used the Nester expression for the  
ADM mass in the last line
(when appropriately generalized to five dimensions \cite{Sparling}). 
Evaluating $D_\nu E^{\mu\nu}$ we get
$D_\nu E^{\mu\nu}=\frac{1}{2}(\overline{D_\nu\epsilon})\gamma^{\mu\nu\rho}
D_\rho\epsilon +c.c.$
such that
\begin{equation}
 {\cal M}=
\int_{\partial\Sigma} dS_{\mu}
(\frac{1}{2}(\overline{D_\nu\epsilon})
\gamma^{\mu\nu\rho}D_\rho\epsilon +c.c.)
\end{equation}
Provided that $\epsilon$ satisfies the modified Witten condition 
$n^\mu D_\mu\epsilon =0$,
where the vector 
$\vec{n}$  is normal to $\Sigma$,
this integral is non-negative \cite{Hull}. Hence, we have the general result
$\overline{\epsilon}_\infty \gamma^\mu\epsilon_\infty\, ^5P_\mu ^{ADM}\geq 0$, 
where $\epsilon_\infty$ is the value of the spinor near $\partial\Sigma$.
The bound is in our case saturated,
and the mass is minimized,
 when $\epsilon$ is a Killing spinor, i.e. satisfies $D_\mu\epsilon =0$,
as can be seen directly from the equations above.   

We now turn to the question of whether the GPS solution can carry a spin structure of
the Killing type. The condition to be satisfied is explicitly given by
\begin{equation}
D_\mu\epsilon =(\partial_\mu +\frac{1}{4}
\omega_\mu\, ^{BC}\gamma_{BC})\epsilon =0
\end{equation}
It is clear that this condition will imply particular relations between
the four dimensional gravitational field, the
dilaton and the four dimensional electro-magnetic field $F$.
A basic assumption which is realized in the GPS solution is
that the f\"{u}nf-bein is independent of $x^5$. This property of the
f\"{u}nf-bein is in general inherited by the gravitini and the local SUSY transformation
parameters due to the SUSY transformations
eq.(5). Due to the existence of this isometric direction
we will find it sufficient to only consider 
the fifth component ($\mu$=5) of the (now purely geometric) Killing spinor equation
eq.(9).
It follows that this equation will induce an algebraic relation between
certain spin coefficients since $\partial_5\epsilon =0$. Explicitly we get
the general result that
$\frac{1}{8}\omega_5^{BC}\gamma_{BC}\epsilon =0$,
or when ``projected'' down into four dimensions
\begin{equation}
(2(\partial_je^{-\frac{\phi}{\sqrt{3}}})\gamma^{j5}-
e^{-\frac{2\phi}{\sqrt{3}}}F_{ij}\gamma^{ij})
\epsilon =0
\end{equation}
where the $5$-index in eq.(10) is now a local f\"{u}nf-bein index.
This last relation is extremely restrictive as to which 
kind of ``physical'' field combinations
that can be supported by the theory in four dimensions.
Consider the general (hypothetical) class of 
dyonic four dimensional static and spherically symmetric monopole
configurations with both a radially directed electric field, as well as
a radially directed magnetic field.
In order to solve eq.(10) for such configurations we consider
this equation as being part of the
 eigen-value problem for the matrix multiplying the
spinor $\epsilon$.
The zero eigen-values will solve eq.(10).
 It is easy to see that the eigen-values
of the matrix in
eq.(10) in general will be complex. The eigen-values are zero only
provided that all the charges in the system
are put identically to zero. 
The GPS solution is of the above type, since it displays a radially directed
magnetic field, something one easely deduces from the 
associated five-geometry, which can be written in the form \cite{Gross,Sorkin1}
\begin{eqnarray}
ds^2=-dt^2+V(dx^5+Ad\phi )^2+V^{-1}(dr^2+r^2d\Omega_2^2)\, ,\\
V\equiv (1-\frac{4m}{r})^{-1}\,\, ,\,\, A\equiv 4m(1-\cos\theta )\, ,
\end{eqnarray}
with the aid of the explicit relations in eq.(3) between the
f\"{u}nf-bein and the vier-bein fields. $m$ is a mass parameter, 
$\phi$ parameterizes the usual periodic
space like Killing trajectory in
spherical polar coordinates equipped with the standard infinitesimal line element
$d\Omega_2^2$ on the unit sphere.
Hence, the GPS monopole can 
in particular not support a non-trivial
spin structure of the Killing type.

Our findings run contrary to previous expectations concerning
the stability of the GPS solution. Stability of the monopole was argued for
in \cite{Sorkin2}, since the ADM mass
has been argued to equal the magnetic charge of the solution. 
Hence, the
solution seemingly also saturates a Bogomol'nyi type bound of the kind eq.(1),
which was also derived in \cite{Sorkin2}. 
However, as we have shown above,
saturation of this bound implies the
existence of a Killing spinor field, 
something which was tactically assumed to
be present in \cite{Sorkin2}. Our results invalidate this basic assumption.

How can the absence of a Killing spinor be reconciled with the result that
apparently $M_{ADM}=Q_M$ ? A fundamental problem in deriving this last equality
for the GPS solution is that the four geometry exhibits a naked
singularity. Solitons in theories without a coupling to the gravitational field
 is {\em defined}
to be non-singular. In any calculational scheme used in order to deduce the
value of the ADM mass of the monopole, the mass of the singularity must also be
accounted for. There does not exist a well defined procedure which can tell us 
how this should be done
in the general situation. In e.g. \cite{Bombelli} this question was 
``bypassed''
by arguing that different calculational schemes
(both five and four dimensional ones) give the same value on the
ADM mass of the monopole
 when the presence of the singularity in the four geometry
 was appropriately ``suppressed''. Our
findings indicate that the singular nature of the GPS solution was not
properly taken into account in these calculations, and  we
consequently conclude that $M_{ADM}\neq Q_M$. This implies that the
GPS monopole is not necessarily stable. This conclusion is in line with
what one would naively expect, namely that a theory of the Einstein kind
cannot support structures of the BPS type. However, we are not
able to definitely conclude that the GPS solution is unstable, since the
significance of the $R^4\times S^1$ type topology of the 
five dimensional manifold for the stability
issue is generally somewhat unclear \cite{Witten2}.

\smallskip

{\bf Acknowledgments}
B.J. thanks U.Lindstr\"om and ITP,
University of Stockholm for hospitality during 
the spring semester 1996 where this
work was initiated.  
B.J. also acknowledges partial support from NFR
for a traveling grant 
under project number 110945/432.
Both authors thank U.Lindstr\"om for 
interesting conversations concerning
the subject matter of this paper.

\end{document}